\documentclass[reprint,pra,longbibliography,superscriptaddress]{revtex4-1}
\usepackage{lipsum}
\usepackage{fancyhdr}
\usepackage{graphicx}
\usepackage{braket}
\usepackage[dvipsnames]{xcolor}
\usepackage{mathtools}
\usepackage{amsmath}
\usepackage{amssymb}
\usepackage{hyperref}
\usepackage{esvect}
\usepackage{float}
\graphicspath{{./Figures/}}

\usepackage[colorinlistoftodos,prependcaption]{todonotes}

\usepackage{xargs}
\newcommandx{\greencom}[2][1=]
{\todo[inline, color=green!40,#1]{#2}}
\newcommandx{\bluecom}[2][1=]
{\todo[inline, color=blue!40,#1]{#2}}
\newcommandx{\bluemargin}[2][1=]
{\todo[color=blue!40,#1]{#2}}

\allowdisplaybreaks

\begin{document}

\title{Quantum trajectory theory of few photon cavity-QED systems with a time-delayed coherent feedback}

\author{Gavin Crowder}
\affiliation{Department of Physics, Queen's University, Kingston, Ontario, Canada, K7L 3N6}
\email{14gbc@queensu.ca}
\author{Howard Carmichael}
\affiliation{The Dodd-Walls Centre for Photonic and Quantum Technologies, Department of Physics, University of Auckland, Private Bag 92019, Auckland, New Zealand}
\author{Stephen Hughes}
\affiliation{Department of Physics, Queen's University, Kingston, Ontario, Canada, K7L 3N6}

\begin{abstract}
We describe an efficient approach to modelling  cavity quantum electrodynamics (QED) with a time-delayed coherent feedback using quantum trajectory simulations. An analytical set of equations is derived to exploit the advantages of trajectories in the presence of the non-Markovian dynamics, where adjustments to the standard stochastic dynamics are discussed. In the weak excitation regime, we first verify that our approach recovers known results obtained with other simulation methods and  demonstrate how a coherent feedback loop can increase the photon lifetime in typical cavity-QED systems. We then explore the nonlinear few-photon regime of cavity-QED, under the restriction of at most one photon at a time in the feedback loop. In particular, we show how feedback affects the cavity photoluminescence (populations versus laser detuning), and describe how one must account for conditioning in the presence of feedback, specifically the system observables must be conditioned on no photon detections at the feedback output channel occurring.
\end{abstract}
\maketitle

\section{Introduction}
\label{sec:Intro}

Cavity quantum electrodynamics (cavity-QED), where two-level systems (quantum bits or qubits) are strongly coupled to optical cavities, has been studied in many works, both theoretically \cite{JaynesCummings63,HughesCarmichael11,Carmele13} and experimentally \cite{Ulrich11,Albert11,Nomura10}, with emerging experiments also using qubits embedded in integrated semiconductor microcavities \cite{Lodahl15} or implemented in super-conducting circuits (circuit-QED) \cite{Gu17}. These elementary quantum systems often couple to integrated waveguides to give greater ``on-chip'' control, emitting single photons into a waveguide mode \cite{Hoang16}, even in only one direction -- ``chiral waveguides''\cite{Coles16,ChiralPaper}. Cavity-QED systems can also aid quantum information objectives in other ways, e.g., by generating squeezed light \cite{Wiseman06}. Due to their often short photon lifetimes, however, they can lack the long-term stability required by applications.

Quantum feedback has been proposed as a method to increase stability (and coherent lifetime) in cavity-QED, by coupling the system to an optical feedback loop that coherently returns photons after a time delay \cite{Lloyd00}. Other suggested applications of coherent feedback include stabilization in optomechanics \cite{Naumann16}, enhancement of photon entanglement \cite{Hein16} and squeezing \cite{Kraft16,Nikolett16}, enhanced photon bunching/antibunching and improved photon distribution in quantum emitters \cite{Lu17,Droenner19}.

The most commonly used approach to solve for the evolution of cavity-QED systems employs open-system quantum master equations \cite{QuantumNoise}, which readily include dissipation by tracing over the reservoir \cite{Carmichael02}. Time-delayed coherent feedback, however, contradicts one of the fundamental assumptions of the approach, namely the assumption of a Markovian dynamic. Under the Markov approximation, the system evolution must depend on the present system state only (local time) and not draw on information from the past. Time-delayed feedback explicitly violates this requirement and so a new approach is called for.

Due to complexities arising from the continuum of modes in the feedback reservoir and the non-Markovian dynamic, the majority of studies have been limited to the linear regime; although nonlinearity at the few-quanta level has been treated by employing fictitious cascading systems \cite{Grimsmo15,Whalen17} or matrix product states \cite{MatrixProductStates,Pichler16,Droenner19,Guimond17}. These treatments focus on particular model systems, however, and they can meet with severe computational limitations as photon numbers increase. There is thus reason to develop alternative approaches to the modeling of time-delayed coherent feedback in quantum optics, especially ones that offer more physical insight into the underlying dynamics.

In this work, we introduce  an intuitive approach that exploits the physical insight and numerical efficiency provided by quantum trajectory (QT) simulations. It is naturally suited to the coherent feedback problem, allowing us to incorporate the non-Markovian effects while preserving the usual benefits of a QT evolution -- in particular, the linear scaling with the overall size of the Hilbert space, which brings distinct advantages in the multiphoton regime. 

\begin{figure}[tbp]
\includegraphics[width=0.499\textwidth]{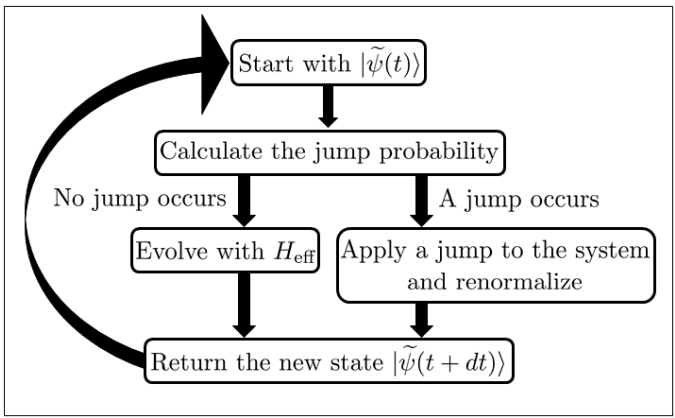}
  \caption{\footnotesize{Schematic representation of one time step in the quantum trajectory evolution.}}
  \label{Timestep}
\end{figure}

A QT evolves the (not necessarily normalized) ket vector, $\ket{\widetilde{\psi} (t)}$, of a cavity-QED system \cite{Dalibard92,Molmer93,Carmichael92} according to the nonunitary Schr\"{o}dinger equation
\begin{equation}
\frac{d}{dt} \ket{\widetilde{\psi} (t)} = -i H_{\rm{eff}} \ket{\widetilde{\psi} (t)},
\label{BetweenJump}
\end{equation}
where $H_{\rm{eff}}$ is the effective non-Hermitian  Hamiltonian and we adopt natural units with $\hbar = 1$. This evolution is augmented by quantum jumps, at random times, which account for the dissipation operators, $\{ C_i \}$, in the master equation; thus at each time step, the integrated probability for the jump $C_j$ to occur is given by
\begin{equation}
P_j(t) = \int_{t_0}^t \delta p_j(t') dt',
\end{equation}
where $t_0$ is the time of the last jump (any $C_j$) and $\delta p_j(t') = \delta t \braket{\widetilde{\psi}(t')|C_j^{\dagger} C_j|\widetilde{\psi}(t')}$; and $\delta t$, the size of the time step, is assumed sufficiently small that the probability for two jumps to occur in a time step may be neglected \cite{Dalibard92}. A schematic representation of the evolution for one time step is shown in Fig.~\ref{Timestep}. The average over many QTs can be shown to recover the evolution of the density operator of the open system \cite{Molmer93} (if desired), and single trajectories provide unique insight into the underlying stochastic dynamics. Furthermore, since each trajectory is independent of the others, numerical computations can be parallelized.

The rest of our paper is organized as follows. In Sec.~\ref{sec:Model}, we present the main model of interest and show how the feedback loop is added to the QT formalism and treated in the Hamiltonian; a discussion of the ``good" and ``bad" cavity regimes is also presented. In Sec.~\ref{sec:QTT}, we implement the proposed approach and spell out the key analytical steps that are taken to make the evolution under $H_{\rm{eff}}$ tractable for this system; we also show how to evaluate the probability of a quantum jump occurring and how to execute a quantum jump. To enable the computation  of system observable expectations (e.g., the mean cavity photon number), we also discuss the need for conditioning in the presence of feedback, and present a numerical solution to this problem. We then discuss the numerical algorithm built to sample QTs in Sec.~\ref{sec:Numerics}, and, lastly, in Sec.~\ref{sec:Results}, run through a variety of results from simulations made in different parameter regimes: firstly, previous results are replicated to confirm the accuracy of the treatment, and coherent feedback is shown to increase photon lifetimes; multiphoton effects (beyond weak excitation) are then discussed and examples of possible improvements arising from coherent feedback are explored. Our conclusions are presented in Sec.~\ref{sec:Conc}, and some details from Sec.~\ref{sec:QTT} are moved to Appendices~\ref{sec:App1} and \ref{sec:App2}. An optimized QT technique for a simplified version of the model is presented in Appendix~\ref{sec:App3}.

\section{Model and Hamiltonian}
\label{sec:Model}

\begin{figure}[tbp]
\includegraphics[width=0.499\textwidth]{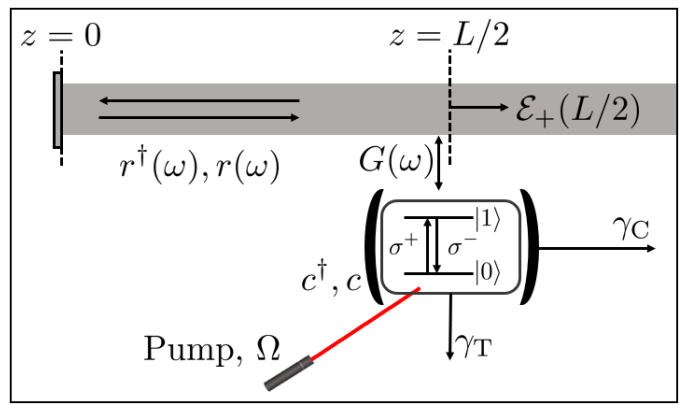}
  \caption{\footnotesize{Schematic of the cavity-TLS coupled to a waveguide at $L/2$. The system has output channels $C_0$ (decay rate $\gamma_{\rm{C}}$), $C_1$ (decay rate $\gamma_{\rm{T}}$), and $\mathcal{E}_+(L/2)$ (the field propagating to the right out of the waveguide), a feedback loop (round-trip length $L$) with a perfect mirror at $z=0$, and is driven by a CW laser of strength $\Omega$. The cavity creation, annihilation operators are $c^{\dagger}, c$; the TLS raising, lowering operators are $\sigma^+, \sigma^-$; and the feedback loop creation, annihilation operators are $r^{\dagger}(\omega), r(\omega)$. The ground and excited states for the TLS are denoted by $\ket{0}$ and $\ket{1}$.}}
  \label{Model}
\end{figure}

The cavity-QED feedback model under investigation comprises a cavity coupled to a two-level system (TLS) as depicted in Fig.~\ref{Model}. The TLS has raising and lowering operators $\sigma^+$ and $\sigma^-$, and the creation and annihilation operators for the cavity mode are $c^\dagger$ and $c$. The system is driven by a continuous-wave (CW) laser of Rabi frequency $\Omega$ and coupled to three output channels: $C_0$ representing cavity decay to an open reservoir (no feedback), $C_1$ representing spontaneous decay from the TLS, and $\mathcal{E}_+(L/2)$ representing the field propagating to the right (and out of the system) at $z=L/2$ in the waveguide. Of particular interest is the coupling to the feedback reservoir, which takes photons emitted from the cavity to the left in the waveguide and feeds them  back {\it coherently}; as shown in Fig.~\ref{Model}, photons entering the feedback loop are reflected by a mirror -- e.g., a microcavity, which could introduce additional loss\cite{HughesCoupledQED} -- and returned to the cavity after a time delay $\tau = {L}/{c(\omega)}$, with $L$ the round-trip length and $c(\omega)$ the speed of photons, of frequency $\omega$, in the feedback loop. The feedback loop is modeled as a continuum of photon modes, with creation and annihilation operators $r^{\dagger}(\omega)$ and $r(\omega)$. In practice, such schemes can be realized, e.g., using quantum dots in photonic waveguide crystal structures \cite{NRCsystems,Lodahl15,HughesCoupledQED}.

The system is investigated in both the ``good" and ``bad" cavity regimes. In the former case, photons have a non-negligible lifetime in the cavity and non-vanishing photon populations accrue; this is usually achieved with $g > \gamma_{\rm{C}}, \gamma_{\rm{L}}$, $\gamma_{\rm{T}}$, where $g$ is the cavity-TLS coupling rate, $\gamma_{\rm{C}}$ and $\gamma_{\rm{T}}$ are the decay rates of the cavity and TLS to the open reservoir, respectively, and $\gamma_{\rm{L}}$ is the decay rate into the waveguide. In the ``bad" cavity case, the photon population in the cavity should quickly decay to zero; however, different choices of parameters achieve this and there is no clear ``bad" cavity response: setting $\gamma_{\rm{C}}, \gamma_{\rm{T}} = 0$ and $\gamma_{\rm{L}} \gg g$, the loop can be engineered such that little population is lost from the overall system (cavity-TLS plus feedback reservoir) and the feedback loop maintains a photon population in the TLS; on the other hand, setting $\gamma_{\rm{L}} = 0$ and $\gamma_{\rm{C}}, \gamma_{\rm{T}} \gg g$, the cavity quickly decays and no photon population is present. Both parameter choices yield a ``bad" cavity, but the dynamic is completely different and so care must be taken in this regime.

Our model is captured by the Hamiltonian
\begin{align}
H ={} & \delta_{aL} \sigma^+\sigma^- + \delta_{cL} c^{\dagger}c + \int_{-\infty}^{\infty}[\omega' r^{\dagger}(\omega') r(\omega')] d\omega' \\
 & \hspace{0.25cm} + g (\sigma^+ c + c^{\dagger}\sigma^-) + \Omega (\sigma^+ + \sigma^-) \nonumber \\
 & \hspace{0.25cm} + \int_{-\infty}^{\infty} [G(\omega')(c^{\dagger}r(\omega') + r^{\dagger}(\omega') c)] d\omega', \nonumber
\end{align}
where $\delta_{aL}=\omega_a-\omega_L$ and $\delta_{cL}=\omega_c-\omega_L$ are the detunings of the TLS and cavity, respectively, from the frequency, $\omega_L$, of the laser drive, and $G(\omega) = \sqrt{\gamma_{\rm{L}}/2\pi} \sin [(\omega \tau + \phi)/2]$\citep{Carmele13,Nemet19} is the frequency dependent coupling between the cavity and the feedback reservoir, where $\phi$ is the overall phase change around the feedback loop. The Hamiltonian is written in an interaction picture designed to remove the oscillation at the frequency of the drive, and we have made a rotating wave approximation.

\section{Quantum Trajectory Theory}
\label{sec:QTT}

In QT theory~\cite{Dalibard92,Molmer93,Carmichael92}, when no quantum jump occurs, the system evolves coherently according to \eqref{BetweenJump}, where the non-Hermitian Hamiltonian, $ H_{\rm eff} = H - \frac{i}{2} (C_0{^\dagger} C_0 + C_1{^\dagger} C_1) $, is the system Hamiltonian augmented by the Lindblad jump operators that represent cavity decay to the open reservoir and spontaneous emission. If a quantum jump occurs, then one of the jump operators, $C_0$, $C_1$, or $\mathcal{E}_+(L/2)$, is applied to the system and the state is renormalized. The challenge in applying this method to our model arises in the algebraic form of the system state, which must encompass both the cavity-TLS and the feedback loop:
\begin{align}
\ket{\psi (t)} ={} & \sum_{n=0}^{N} \left[\vphantom{\int} \left( \alpha_n(t)\ket{0} + \beta_n(t) \ket{1} \right) \ket{ \{ 0 \} } \right.
\label{SystemModel}\\
 &\mkern-60mu + \left. \int_{-\infty}^{\infty} \left( R_{0,n}(\omega',t)\ket{0} + R_{1,n}(\omega',t)\ket{1} \right) \ket{1,\omega'} d\omega' \right]\mkern-2mu \ket{n}, \nonumber
\end{align}
where we restrict the expansion to just one photon in the feedback loop, though many photons may occupy the cavity (the index $n$); $\alpha$ and $\beta$ represent the amplitudes of the ground and excited state, respectively, of the TLS when there is no photon in the feedback loop, while $R_0$ and $R_1$ represent those amplitudes with one photon in the feedback loop; the frequency of the photon in the feedback loop is indicated by $\omega'$, and the notation $\{ 0 \}$ indicates no photon in every mode of the loop, i.e., the vacuum of the feedback loop. The restriction to at most one photon in the feedback loop at any time assumes either $\gamma_{\rm{L}} \tau \ll 1$ or $\Omega \ll 1$; it neglects multiphoton interference effects, which will be incorporated in future work. This is expected to be a reasonable approximation for small feedback loops and low loss.

Since there is a continuum of reservoir modes, $R_{0,n}(\omega,t)$ or $R_{1,n}(\omega,t)$ cannot be evolved individually, or at least it would be numerically cumbersome to do so. For a more efficient approach, we first find an expression for these amplitudes in terms of $\alpha_n(t)$ and $\beta_n(t)$, with the number of photons in the cavity truncated at $n_{\rm{max}} = N$.

Substituting the expansion of \eqref{SystemModel} into  \eqref{BetweenJump}, and employing the above expression for $H_{\rm{eff}}$ with $C_0 = \sqrt{\gamma_{\rm{C}}} a$ and $C_1 = \sqrt{\gamma_{\rm{T}}} \sigma^-$, we arrive at a set of $4(N+1)$ coupled differential equations:
\begin{align}
\frac {d\alpha_n}{dt} ={} & - A_n\alpha_n - ig\sqrt{n} \beta_{n-1} \label{FullSet} \\
 & \hspace{-0.25cm}\mkern14mu -i\Omega \beta_{n} - i \int_{-\infty}^{\infty} G(\omega') \sqrt{n} R_{0,n-1} (\omega') d\omega', \nonumber \\
\frac {d\beta_n}{dt} ={} & -B_n\beta_n - ig\sqrt{n+1} \alpha_{n+1} \nonumber \\
 & \hspace{-0.25cm}\mkern14mu  -i\Omega \alpha_{n} - i \int_{-\infty}^{\infty} G(\omega') \sqrt{n} R_{1,n-1} (\omega') d\omega', \nonumber \\
\frac {dR_{0,n}(\omega)}{dt} ={} & -(A_n + i\omega) R_{0,n}(\omega) - ig\sqrt{n} R_{1,n-1}(\omega) \nonumber \\
 & \hspace{-0.25cm}\mkern14mu -i\Omega R_{1,n}(\omega) - i G (\omega) \sqrt{n+1} \alpha_{n+1},{} \nonumber \\
\frac {dR_{1,n}(\omega)}{dt} ={} & -(B_n + i\omega) R_{1,n}(\omega) - ig\sqrt{n+1} R_{0,n+1}(\omega) \nonumber \\
 & \hspace{-0.25cm}\mkern14mu -i\Omega R_{0,n}(\omega) - i G (\omega) \sqrt{n+1} \beta_{n+1}, \nonumber
\end{align}
where $A_n = n\gamma_{\rm{C}}/2 + i n \delta_{cL}$ (note $A_0 = 0$) and $B_n = (\gamma_{\rm{T}} + n\gamma_{\rm{C}})/2 + i (\delta_{aL} + n\delta_{cL})$; amplitudes indexed by $n$ with $n \notin [0,N]$ are zero, e.g., $R_{0,N+1}(\omega,t) = 0$ at all times.

Keeping our goal in mind, i.e., to obtain a set of easy to evolve equations for $\alpha_n(t)$ and $\beta_n(t)$, our task now is to obtain suitable expressions for $R_{0,n-1} (\omega ,t)$ and $R_{1,n-1} (\omega ,t)$ for substitution on the right-hand sides of the first two state amplitude equations, \eqref{FullSet}; with the integrals over $d\omega'$ evaluated, we aim for a closed set of coupled differential equations. We show here how to proceed in the $N = 1$ case, with the general $N$ case following the same approach but with more algebraic complexity. This allows for up to two quanta in the cavity-TLS (the cavity and TLS both excited) and goes beyond the one-quantum treatments commonly encountered in the literature (e.g., see Ref.~\onlinecite{Carmele13}).

\subsection{Decoupling $R_{j,n}$ in the $N=1$ Case}
\label{sec:Decoupling}

With the maximum number of photon states in the cavity set to $N = 1$, the coupled differential equations for $R_{j,n}(\omega ,t)$ reduce to
\begin{align}
\frac {dR_{0,0}(\omega)}{dt} ={} & -i \omega R_{0,0} (\omega) - i \Omega R_{1,0}(\omega) - i G(\omega) \alpha_1, \\
\frac {dR_{1,0}(\omega)}{dt} ={} & -(B_0 + i\omega) R_{1,0} (\omega) - i \Omega R_{0,0}(\omega) \nonumber \\
& \hspace{0.5cm} - i g R_{0,1}(\omega) - i G(\omega) \beta_1, \nonumber \\
\frac {dR_{0,1}(\omega)}{dt} ={} & - (A_1 + i\omega) R_{0,1} (\omega) - i \Omega R_{1,1}(\omega) \nonumber \\
& \hspace{0.5cm} - i g R_{1,0}(\omega), \nonumber \\
\frac {dR_{1,1}(\omega)}{dt} ={} & -(B_1 + i\omega) R_{1,1} (\omega) - i \Omega R_{0,1}(\omega). \nonumber
\label{CoupledEq}
\end{align}
If we then define two vectors,
\begin{equation}
\boldsymbol{R}(\omega,t) \equiv \begin{pmatrix}
           R_{0,0} (\omega ,t)\\           
           R_{1,0} (\omega ,t)\\
           R_{0,1} (\omega ,t)\\
           R_{1,1} (\omega ,t)
          \end{pmatrix}, \hspace{0.25cm}
\boldsymbol{\alpha\beta}(t) \equiv \begin{pmatrix}
           \alpha_1 (t) \\
           \beta_1 (t) \\
           0 \\
           0
          \end{pmatrix},
\end{equation}
and the matrix
\begin{equation}
\boldsymbol{A} \equiv \begin{bmatrix}
-i \omega & -i \Omega & 0 & 0\\           
-i \Omega & -(B_0 + i\omega) & -ig & 0\\
0 & -ig & -(A_1 + i\omega) & -i\Omega \\
0 & 0 & -i\Omega &  -(B_1 + i\omega)
\end{bmatrix},
\end{equation}
this system may be written in the simple form
\begin{equation}
\frac{d}{dt} \boldsymbol{R}(\omega,t) = \boldsymbol{A} \boldsymbol{R}(\omega,t) -iG(\omega) \boldsymbol{\alpha\beta}(t),
\label{DiffEQ}
\end{equation}
with solution
\begin{equation}
\boldsymbol{R} (\omega,t) = -i G(\omega) \int_0^t \boldsymbol{E} e^{-\boldsymbol{\lambda}(t' -t)} \boldsymbol{E}^{-1} \boldsymbol{\alpha\beta}(t') dt',
\end{equation}
where $\boldsymbol{E}$ is a matrix formed from the eigenvectors of $\boldsymbol{A}$ and $\boldsymbol{\lambda}$ is a diagonal matrix of eigenvalues. Clearly, the eigenvalues take the form $\lambda_j = -i \omega + c_j$, $c_j \in \mathbb{C}$. Thus, we may write
\begin{equation}
\boldsymbol{R} (\omega,t) = -i G(\omega) \int_0^t e^{i\omega(t'-t)} \boldsymbol{E}\cdot \boldsymbol{n}(t,t') dt',
\label{FormOfR}
\end{equation}
where $\boldsymbol{E}$ and $\boldsymbol{n}(t,t')$ are frequency independent. Further details are provided in Appendix \ref{sec:App1}, where the explicit expression for $\boldsymbol{n}(t,t')$ appears as \eqref{nExpression}.

\subsection{Solving the Dynamical Evolution Equations in the $N = 1$ Case}
\label{sec:alphabeta}

Often one is interested in just a few quanta, especially for low loss and good cavity systems; in this case we may restrict the Hilbert space to $N = 1$. This recovers all of the physics in the weak excitation regime -- up to one quantum in the cavity-TLS plus feedback loop -- and, in addition, some multiphoton effects, as we demonstrate in Sec.~\ref{sec:NonlinearResults}. The coupled differential equations for the $\alpha$ and $\beta$ amplitudes are now
\begin{align}
\frac {d\alpha_0}{dt} & = -i \Omega \beta_0, \label{CoupledEq2} \\
\frac {d\beta_0}{dt} &= -B_0 \beta_0 - i \Omega \alpha_0 - i g \alpha_1, \nonumber \\
\frac {d\alpha_1}{dt} &= - A_1 \alpha_1 -i \Omega \beta_1 - i g \beta_0 - i \int_{-\infty}^{\infty} G(\omega') R_{0,0} (\omega') d\omega',{} \nonumber \\
\frac {d\beta_1}{dt} & = - B_1 \beta_1 -i \Omega \alpha_1 - i \int_{-\infty}^{\infty} G(\omega') R_{1,0} (\omega') d\omega', \nonumber
\end{align}
where the coupling to the feedback loop is made through the quantities \eqref{FormOfR}:
\begin{align}
R_{0,0}(\omega,t) ={} & -i G(\omega) \int_0^t e^{i\omega(t'-t) }\boldsymbol{E}_1 \cdot \boldsymbol{n}(t,t') dt', \\
R_{1,0}(\omega,t) ={} & -i G(\omega) \int_0^t e^{i\omega(t'-t) }\boldsymbol{E}_2 \cdot \boldsymbol{n}(t,t') dt', \nonumber
\end{align}
where $\boldsymbol{E}_1$ and $\boldsymbol{E}_2$ are the first and second rows of $\boldsymbol{E}$, respectively. As is shown in Appendix \ref{sec:App2}, the required double integrals simplify to give
\begin{align}
-i \int_{-\infty}^{\infty} G(\omega') R_{0,0} (\omega',t) d\omega' ={} & \label{RIntegral1} \\
 & \hspace{-3.7cm} \frac{\gamma_{\rm{L}}}{4} \left[ -\boldsymbol{E}_1 \cdot \boldsymbol{n} (t,t) + e^{i\phi}\theta(t - \tau) \boldsymbol{E}_1 \cdot \boldsymbol{n} (t,t-\tau) \right], \nonumber
\end{align}
and
\begin{align}
-i \int_{-\infty}^{\infty} G(\omega') R_{1,0} (\omega',t) d\omega' ={} & \label{RIntegral2} \\
 & \hspace{-3.7cm} \frac{\gamma_{\rm{L}}}{4} \left[ -\boldsymbol{E}_2 \cdot \boldsymbol{n} (t,t) + e^{i\phi}\theta(t - \tau) \boldsymbol{E}_2 \cdot \boldsymbol{n} (t,t-\tau) \right], \nonumber
\end{align}
where $\theta(\tau)$ is the Heaviside step function. 

The set of equations for $\alpha$ and $\beta$ amplitudes is now closed and can be used to evolve the state of the entire system, \eqref{SystemModel} with $N = 1$, between quantum jumps. It is worth noting that $\boldsymbol{R} (\omega,t)$ can be determined at any time ($t$) and frequency ($\omega$) as it only depends on the history of $\alpha(t)$ and $\beta(t)$, which is known. Thus, expectations relating to the feedback loop can be calculated, such as the photon population in the loop.

\subsection{Calculating the Probability for a Quantum Jump}
\label{sec:JumpProbability}

The evolution between jumps has been solved, and now the probability of a quantum jump occurring during a time step $\delta t$ must be determined. This is given by a sum of the expectation values:
\begin{equation}
\delta p (t) = \delta t \Braket{\widetilde{\psi}(t) | \mathcal{E}^{\dagger}_+(L/2) \mathcal{E}_+(L/2) + \sum_{i=0}^1 C_i^{\dagger}C_i | \widetilde{\psi}(t)},
\end{equation}
where $C_0 = \sqrt{\gamma_{\rm{C}}}a$, $C_1 = \sqrt{\gamma_{\rm{T}}}\sigma^-$, and $\mathcal{E}_{+} (L/2) = \frac{-i}{\sqrt{2\pi}}  \int_{-\infty}^{\infty} e^{i \omega \tau /2} r(\omega) d\omega$. Only the evaluation of the $C_0^\dagger C_0$ expectation is shown as all three are evaluated in a similar way. We have
\begin{align}
\braket{\widetilde{\psi}(t) | C_0^{\dagger} C_0 |\widetilde{\psi}(t)} ={} & \gamma_{\rm{C}} \left[ |\alpha_1(t)|^2 + |\beta_1(t)|^2 \right. \\
 & \hspace{-2cm} + \left. \int_{-\infty}^{\infty} |R_{0,1}(\omega',t)|^2 + |R_{1,1}(\omega',t)|^2 d\omega' \right], \nonumber
 \label{Expectation}
\end{align}
where, from \eqref{FormOfR}:
\begin{align}
R_{0,1}(\omega,t) ={} & -i G(\omega) \int_0^t e^{i\omega(t'-t)} \boldsymbol{E}_3 \cdot \boldsymbol{n}(t,t') dt', \\
R_{1,1}(\omega,t) ={} & -i G(\omega) \int_0^t e^{i\omega(t'-t)} \boldsymbol{E}_4 \cdot \boldsymbol{n}(t,t') dt'. \nonumber
\end{align}
Thus, we need the integral
$$
I_{\mu} = \int_{-\infty}^{\infty} R^*_{\mu-3,1}(\omega',t) R_{\mu-3,1}(\omega',t) d\omega',
$$
where $\mu = 3$ or $4$, which expands as 
\begin{align}
I_{\mu} ={} & \int_0^t \int_0^t [\boldsymbol{E}_{\mu} \cdot \boldsymbol{n}(t,t'')]^* [\boldsymbol{E}_{\mu} \cdot \boldsymbol{n}(t,t')] \\
& \hspace{-0cm} \times \int_{-\infty}^{\infty} G^2(\omega') e^{-i\omega'(t''-t) } e^{i\omega'(t'-t) } d\omega' dt' dt''. \nonumber
\end{align}
Substituting the exponential form of $G(\omega')$, we then carry out the integration over frequency to arrive at
\begin{align}
I_{\mu} ={} & \frac{\gamma_{\rm{L}}}{4} \int_0^t \int_0^t  dt' dt'' [\boldsymbol{E}_{\mu} \cdot \boldsymbol{n}(t,t'')]^* [\boldsymbol{E}_{\mu} \cdot \boldsymbol{n}(t,t')] \\
& \hspace{-0.8cm}\mkern-5mu \times \left[ 2\delta(t'' - t') - e^{i\phi} \delta(t'' -\tau -t') - e^{-i\phi} \delta(t'' + \tau -t') \right]. \nonumber
\end{align}
Finally, the integral with respect to $t'$ is carried out to yield a sum of three terms:
\begin{align}
I_{\mu} &= \frac{\gamma_{\rm{L}}}{4} \left[ \int_0^t 2 |\boldsymbol{E}_{\mu} \cdot \boldsymbol{n}(t,t'')|^2 dt''  \right. \\
& \hspace{0.5cm} - e^{i\phi} \int_{\tau}^t (\boldsymbol{E}_{\mu} \cdot \boldsymbol{n}(t,t''))^* (\boldsymbol{E}_{\mu} \cdot \boldsymbol{n}(t,t''-\tau)) dt'' \nonumber \\
&\hspace{0.5cm} \left. - e^{-i\phi} \int_{0}^{t-\tau} (\boldsymbol{E}_{\mu} \cdot \boldsymbol{n}(t,t''))^* (\boldsymbol{E}_{\mu} \cdot \boldsymbol{n}(t,t''+\tau)) dt'' \right]. \nonumber
\end{align}
It is readily shown that the second and third terms are complex conjugates of one another so there are just two integrals to compute in a numerical implementation. From \eqref{Expectation} and the similar result for the expectation of $C_1^{\dagger} C_1$, we now have:
\begin{align}
\braket{\widetilde{\psi}(t)| C_0^{\dagger} C_0 | \widetilde{\psi}(t)} ={} & \gamma_{\rm{C}} \left[ |\alpha_1(t)|^2 + |\beta_1(t)|^2 + I_3 + I_4 \right], \\
\braket{\widetilde{\psi}(t) | C_1^{\dagger} C_1 | \widetilde{\psi}(t)} ={} & \gamma_{\rm{T}} \left[ |\beta_0(t)|^2 + |\beta_1(t)|^2 + I_2 + I_4 \right]. \nonumber
\end{align}

Although derived in a similar way, the expectation of $\mathcal{E}^{\dagger}_+(L/2) \mathcal{E}_+(L/2)$ does not require integrals over the past time and is 
\begin{align}
\braket{\widetilde{\psi}(t) | \mathcal{E}_+^{\dagger}(L/2) \mathcal{E}_+(L/2) | \widetilde{\psi}(t)} ={} & \\
& \hspace{-4.2cm} \frac{\gamma_{\rm{L}}}{4} \sum_{i=1}^4 \left| \mathbf{E}_i \cdot \mathbf{n}(t,t) - e^{i\phi} \theta(t-\tau) \mathbf{E}_i \cdot \mathbf{n}(t,t - \tau) \right|^2. \nonumber
\end{align}

\subsection{Applying the Quantum Jump Operator}
\label{sec:ApplyingJumps}

Once a jump is determined to occur and the type of jump chosen using the relative probabilities, one of the jump operators is applied to the system state. For the purposes of illustration, let us assume that a cavity jump occurred and $C_0 = \sqrt{\gamma_{\rm{C}}}a$ operates on the system state. The new \textit{unnormalized} ket vector, at time $t_0$, following the time step in which the jump occurred, is
\begin{align}
\ket{\widetilde{\psi} (t_0)} ={} &  \sqrt{\gamma_{\rm{C}}} \left( [\alpha_1(\bar{t}_0) \ket{0} + \beta_1(\bar{t}_0) \ket{1}] \ket{ \{ 0 \} } \right. \\
& \hspace{-1.2cm}\mkern-7mu \left. + \int_{-\infty}^{\infty} [R_{0,1}(\omega',\bar{t}_0) \ket{0} +R_{1,1}(\omega',\bar{t}_0) \ket{1}] \ket{1,\omega'} d\omega' \right) \ket{0}. \nonumber
\end{align}
where $\bar{t}_0 \equiv t_0 - \delta t$; thus the new non-zero amplitudes are:
\begin{align}
\alpha_0(t_0) & = \sqrt{\gamma_{\rm{C}}} \alpha_1(\bar{t}_0), \\
\beta_0(t_0) & = \sqrt{\gamma_{\rm{C}}} \beta_1(\bar{t}_0), \nonumber \\
R_{0,0}(\omega,t_0) & = \sqrt{\gamma_{\rm{C}}} R_{0,1}(\omega,\bar{t}_0), \nonumber \\
R_{1,0}(\omega,t_0) & = \sqrt{\gamma_{\rm{C}}} R_{1,1}(\omega,\bar{t}_0), \nonumber
\end{align}
while, furthermore, for $t > t_0$, the amplitudes $R_{i,j}(\omega,t)$ are given by:
\begin{align}
i G^{-1}(\omega)R_{0,0}(\omega,t) & = \int_{t_0}^t e^{i\omega(t'-t)} \boldsymbol{E}_1\cdot \boldsymbol{n}(t,t')dt'\label{R_After_Jump} \\
& \hspace{-1cm} + \sqrt{\gamma_{\rm{C}}} \left[ \int_0^{\bar{t}_0}   e^{i\omega(t'-t)} \boldsymbol{E}_3\cdot \boldsymbol{n}(t,t')dt' \right], \nonumber \\
i G^{-1}(\omega)R_{1,0}(\omega,t) & = \int_{t_0}^t e^{i\omega(t'-t)} \boldsymbol{E}_2\cdot \boldsymbol{n}(t,t')dt' \nonumber \\
& \hspace{-1cm} + \sqrt{\gamma_{\rm{C}}} \left[ \int_0^{\bar{t}_0} e^{i\omega(t'-t)} \boldsymbol{E}_4\cdot \boldsymbol{n}(t,t')dt' \right], \nonumber \\
i G^{-1}(\omega)R_{0,1}(\omega,t) & = \int_{t_0}^t e^{i\omega(t'-t)} \boldsymbol{E}_3\cdot \boldsymbol{n}(t,t')dt', \nonumber \\
i G^{-1}(\omega)R_{1,1}(\omega,t) & = \int_{t_0}^t e^{i\omega(t'-t)} \boldsymbol{E}_4\cdot \boldsymbol{n}(t,t')dt'. \nonumber
\end{align}
Thus, both $R_{0,1}$ and $R_{1,1}$ are reset after the jump, while $R_{0,0}$ and $R_{1,0}$ carry the memory of $R_{0,1}$ and $R_{1,1}$ immediately prior to the jump through their initial values at time $t_0$.

This suggests a convenient strategy for implementing $-i \int_{-\infty}^{\infty} G(\omega') R_{i,0} (\omega',t) d\omega'$ in the evolution of $\alpha$ and $\beta$ following a jump. For the $\tau$ time steps after the jump occurs, the terms $\boldsymbol{E}_1\cdot \boldsymbol{n}(t,t-\tau)$ and $\boldsymbol{E}_2\cdot \boldsymbol{n}(t,t-\tau)$ are replaced by $\boldsymbol{E}_3\cdot \boldsymbol{n}(t,t-\tau)$ and $\boldsymbol{E}_4\cdot \boldsymbol{n}(t,t-\tau)$, respectively. This occurs because after the jump the value of $R_{0,0}$ and $R_{1,0}$ evaluated at $\tau$ time steps in the past lands in the integrals from $0$ to $\bar{t}_0$ in \eqref{R_After_Jump}. Care must also be taken when computing the $I_1$ and $I_2$ integrals to ensure that the non-zero overlap of the time integrals are considered when numerically evaluating these quantities.

When a jump down the waveguide occurs -- i.e. when we apply $\mathcal{E}_+(L/2)$ to the system -- a reset of the memory occurs and the history up to that point can be thrown away. We define $\ket{\widetilde{\psi}_{\rm{click}}(t_0)} = \mathcal{E}_+(L/2) \ket{\widetilde{\psi}(\bar{t}_0)}$ to be the unnormalized state of the system after such a jump, where the label ``click'' is explained in the next section. This state is
\begin{align}
    \ket{\widetilde{\psi}_{\rm{click}}(t_0)} ={} & \frac{-i \sqrt{\gamma_{\rm{L}}}}{2} \sum_{i=1}^2 \sum_{n=1}^2 \left[ \mathbf{E}_{\mu} \cdot \mathbf{n}(\bar{t}_0,\bar{t}_0) \right. \\
    & \hspace{-1.5cm} \left. - e^{i\phi} \theta(\bar{t}_0-\tau) (\mathbf{E}_{\mu} \cdot \mathbf{n}(\bar{t}_0,\bar{t}_0-\tau)) \right] \ket{i} \ket{n} \ket{ \{ 0 \} }, \nonumber
\end{align}
where $\mu = 2(n-1) + i$.

\subsection{Conditioning in the Presence of Feedback}
\label{sec:Conditioning}

Due to the non-Markovian nature of the output channel $\mathcal{E}_{+} (L/2)$, the populations of the system must be conditioned on no jump occurring (denoted ``noclick'') in a different way to typical Markovian output channels such as $C_0$. Rather than introducing a non-unitary part to the system Hamiltonian (since photon loss from the feedback loop is not Lindblad), the conditioning is implemented explicitly by hand. Arguing from the standard theory of photon counting \cite{Carmichael87}, after a jump at time $\bar t_0$, the density operator for the system conditioned on no subsequent jump is
\begin{equation}
    \rho_{\rm{noclick}}(t) = \frac{{\rm{tr}}_{\rm{R}}[\chi_{\rm{noclick}} (t)]}{{\rm{tr}}_{\rm{S \otimes R}}[\chi_{\rm{noclick}} (t)]},
    \label{rhoconditioning}
\end{equation}
where $\chi_{\rm{noclick}} (t) = e^{(\mathcal{L}- \mathcal{E}_+\left(L/2\right)\mkern2mu \cdot\mkern2mu \mathcal{E}_+^{\dagger}\left(L/2\right))(t-\bar t_0)} \chi(\bar t_0)$ with $\mathcal{L} = -i[H_{\rm{eff}},\cdot\mkern3mu]$ and $\chi(\bar t_0) = \ket{\psi(\bar t_0)}\bra{\psi(\bar t_0)}$. Under the approximation of only one photon in the feedback loop at any time, this reduces to
\begin{align}
    \chi_{\rm{noclick}}(t) ={} & \ket{\widetilde{\psi}(t)} \bra{\widetilde{\psi}(t)} \\
    & \hspace{0.5cm} - \int_{\bar t_0}^t e^{\mathcal{L}(t-t')} \ket{\widetilde{\psi}_{\rm click}(t')} \bra{\widetilde{\psi}_{\rm click}(t')} dt'. \nonumber
\end{align}
Note that $\rho_{\rm{noclick}}(t)$ is normalized, so desired observable expectations can be calculated from this quantity.

This process is computationally demanding and can quickly cause the trajectories to take unfeasible amounts of time to run. The interaction with the feedback loop is therefore dropped from $\mathcal{L}$, which is a good approximation under the assumption of at most one photon in the loop -- i.e. $\mathcal{L} \longrightarrow \mathcal{L}_0 = -i[H_{\rm{eff}}^0,\cdot\mkern3mu]$, where 
\begin{align}
    H_{\rm eff}^0 ={} & \delta_{aL} \sigma^+\sigma^- + \delta_{cL} c^{\dagger}c + g (\sigma^+ c + c^{\dagger}\sigma^-) \\
     & \hspace{0.5cm} + \Omega (\sigma^+ + \sigma^-) - \frac{i}{2} \left[ \gamma_{\rm{C}} a^{\dagger} a + \gamma_{\rm{T}} \sigma^+ \sigma^- \right]. \nonumber
\end{align}
After making this approximation, the denominator of \eqref{rhoconditioning} reduces to the probability that no photon has travelled to the right down the waveguide (and left the system) since the last jump. This quantity is given by $1 - P_{\mathcal{E}_+\left(L/2\right)}(t)$, where $P_{\mathcal{E}_+\left(L/2\right)}(t)$ is the integrated probability of jump $\mathcal{E}_+\left(L/2\right)$.

\section{Numerical Implementation}
\label{sec:Numerics}

We implemented this algorithm in Matlab employing the Parallel Computing Toolbox (though it could readily be implemented in other computational languages such as Python). The matrices $\boldsymbol{E}$ and $\boldsymbol{\lambda}$ are first computed from the provided system parameters, and then passed to a parallelized for-loop which runs the QT simulations; since each QT is independent of the others, this calculation is readily parallelized, which leads to a significant saving in computation time. Once the desired ensemble of expectations has been obtained, it is passed back to the main program to be averaged. The approach would work in a similar way with no parallelization, on a single processor, although the computation time would more quickly become prohibitive with increasing complexity of the system.

The QT simulation evolves the provided initial state through enough time steps of sufficient resolution to reach the desired end time. At the start of each time step, the probability, $\delta p$, for a jump to occur is evaluated and the integrated probability, $P(t) = \int_{t_0}^t \delta p(t') dt'$ where $t_0$ is the time of the last jump, is compared against a uniformly distributed random number $\epsilon$: if $\epsilon < P(t)$, a jump is implemented, with the jump operator selected on the basis of a second uniformly distributed random number and the relative jump probabilities from the most recent time step; otherwise the system state is advanced one time step by a modified fourth-order Runge-Kutta algorithm~\cite{Butcher08} (RK4) applied to the equations of motion from Sec.~\ref{sec:alphabeta} -- RK4 makes two evaluations of $\boldsymbol{n} (t,t-\tau + \delta t / 2)$, which is unavailable, and we therefore substitute $\boldsymbol{n} (t,t-\tau)$ for the first evaluation and $\boldsymbol{n}(t,t-\tau + \delta t)$ for the second. Since we are using the integrated probability for a jump at time $t^\prime$ \emph{and} no jump prior to $t^\prime$ (waiting-time distribution), the state is only renormalized after a jump occurs. After the trajectory has completed, the conditioning is done on the system and any desired expectations -- e.g., the population in the TLS, or photon population in the cavity or feedback loop -- can be calculated.

The integrals $I_{\mu}$, $\mu = 1,2,3,4$, must be evaluated once each time step. As they extend over the entire past, they make the largest demand on computation time, which scales quadratically, as a result, with the number of time steps. Since the algorithm scales linearly with the number of QTs, it is more efficient to average many QTs with coarse time resolution than fewer with fine time resolution. An optimized numerical technique for QTs when our system is simplified is presented in Appendix~\ref{sec:App3}.

\section{Results}
\label{sec:Results}

Throughout the results section, we will refer to the TLS population and the cavity population, defined explicitly through the quantities $n_a = \braket{ \psi (t) | \sigma^+ \sigma^- | \psi (t) }$ and $n_c = \braket{ \psi (t) | c^{\dagger} c | \psi (t) }$, respectively.

\subsection{Replication of Previous Results and Quantum Trajectory Insights}
\label{sec:PrevResults}

\begin{figure}[tbp]
  \includegraphics[width=0.499\textwidth]{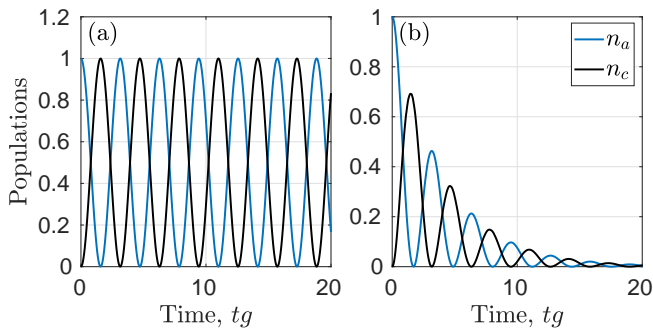}
  \caption{Evolution of the cavity-TLS using the derived QT approach with no feedback loop or drive. The TLS is initially in its excited state so that $\beta_0(t=0) = 1$. In (a), there is no output and vacuum Rabi oscillations occur, while in (b) there is a non-zero output rate ($\gamma_{\rm{L}} = g$) and decaying Rabi oscillations are produced. }
\label{RabiResults}
\end{figure}

To first demonstrate the accuracy of this approach, the numerical model was tested under regimes where the response of the system is already known, or studied elsewhere using different approaches (not QT). Figure~\ref{RabiResults}(a) shows the model as an isolated cavity-TLS system (i.e., $\gamma_{\rm{L}} = \gamma_{\rm{T}} = \gamma_{\rm{C}} = \Omega = 0$) and everything is on resonance, while Fig.~\ref{RabiResults}(b) adds in the waveguide without feedback (i.e., the long loop limit when $\tau \rightarrow \infty$) and with $\gamma_{\rm{L}} = g$. Note that Fig.~\ref{RabiResults}(b) is created by averaging 1000 trajectories, while Fig.~\ref{RabiResults}(a) is created with a single trajectory. This is because without a stochastic decay channel all trajectories will be identical, and so Fig.~\ref{RabiResults}(a) can be created with one trajectory. Since Fig.~\ref{RabiResults}(b) does include a decay channel, an average must be taken in order to recover the ensemble behaviour of the system. These results indeed replicate this relatively simple aspect of cavity-TLS systems~\cite{Ashhab06}.

\begin{figure}[tbp]
  \includegraphics[width=0.499\textwidth]{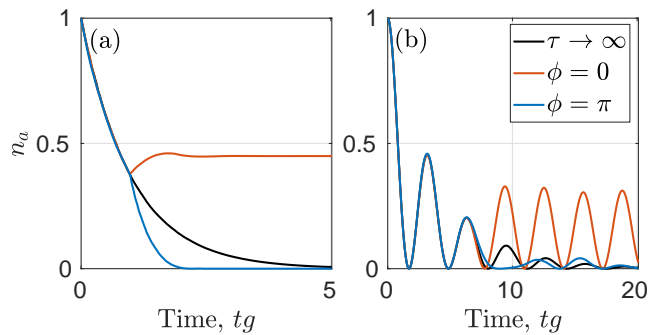}
  \caption{Population dynamics showing the effect of feedback with different $\phi$, when (a) the system  consists of a TLS and (b) the system is a cavity-TLS setup discussed in Sec.~\ref{sec:Model}. Everything is on resonance and there is no drive or Lindblad decay channels; in (a) $\tau = g^{-1}$ and $\gamma_{\rm{L}} = 2g$ while in (b) $\tau = 2\pi g^{-1}$ and $\gamma_{\rm{L}} = g$. Each set of results is an average of 2000 QTs.}
\label{TLSResults}
\end{figure}

Furthermore, the cavity was removed from the model and the TLS coupled directly to the waveguide in order to recover previously studied results \cite{Whalen19,Droenner19}. As shown in Fig.~\ref{TLSResults}(a), by tuning the phase of the feedback loop the system can be made to exhibit enhanced spontaneous emission, when $\phi = \pi$, or trap the excitation, when $\phi = 0$, as long as there are no other loses, $\gamma_{\rm{T}} = \gamma_{\rm{C}} = 0$. Before $t = \tau = g^{-1}$, the time delay introduced by the feedback, the dynamics are identical in all three cases shown as the TLS simply decays. However, as soon as the feedback is first introduced, the dynamics completely change due to interference between the departing and returning fields emitted by the TLS. The same phenomenon can be seen when the cavity is replaced as part of the system, shown in Fig.~\ref{TLSResults}(b), however rather than stabilizing either the TLS or cavity population, the Rabi oscillations are stabilized.

\begin{figure}[tbp]
  \includegraphics[width=0.499\textwidth]{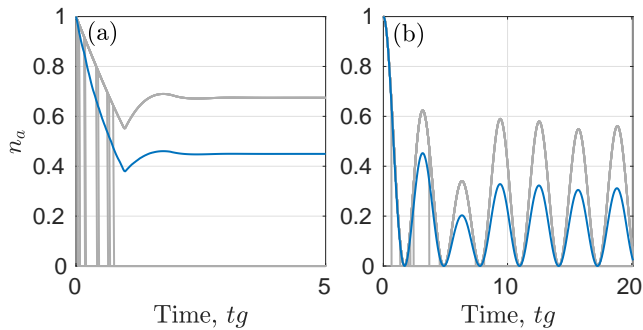}
  \caption{Sample trajectories for the trapping regimes, $\phi = 0$, of Figs.~\ref{TLSResults} (a) and (b). The averaged TLS population is given in blue while the individual trajectories are in grey. Note that if the system does not emit a photon to the right (shown by a jump of $n_a$ to 0) before the feedback returns then the excitation becomes trapped in the system.}
\label{TLSResults_SampleTrajs}
\end{figure}

For both of the trapping regimes presented in Fig.~\ref{TLSResults} there are two types of trajectories that are being averaged together as shown in Figs.~\ref{TLSResults_SampleTrajs}(a) and (b). The individual trajectories for each case are overlaid in grey in the figures; either the system decays and a detector ``click" would occur or the detector does not ``click" and we are left with a trapped excitation. Note that the ``click" only occurs before the feedback has returned from its first round trip, after this time the excitation is trapped if the ``click" has not occurred. The final average, shown in blue, is the average of the trajectories sitting in the ground state and the trajectories with a trapped excitation.

\begin{figure}[tbp]
    \includegraphics[width=0.499\textwidth]{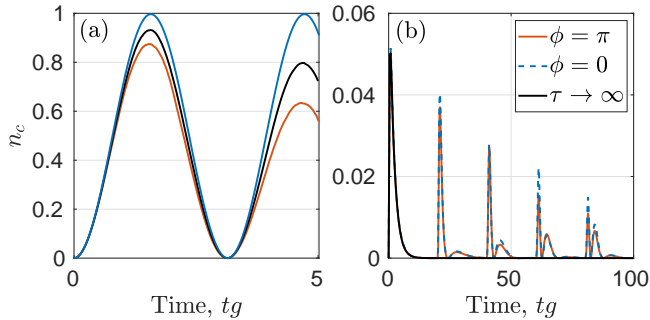}
  \caption{Population dynamics of (a) a short delay time ($\tau = 0.1 g^{-1}$) and (b) a long delay time ($\tau = 20 g^{-1}$). Both sets of results are the average of  2000 QTs with everything on resonance and no drive or Lindblad decay channels. The decay rate into the waveguide is (a) $\gamma_{\rm{L}} = 0.2 g$ and (b) $\gamma_{\rm{L}} = 5\pi g$.}
\label{ShortLong}
\end{figure}

Lastly, the recent results by N\'emet {\it{et al.}} \cite{Nemet19} are also recovered by this approach as shown in Fig.~\ref{ShortLong}. In this setup, the effects of a very short or very long feedback loop were investigated. When the delay time is very small compared to the lifetime of the system, the phase change, $\phi$, from the loop has an immediate and significant effect on the system as shown in Fig.~\ref{ShortLong}(a). Figure~\ref{ShortLong}(b) shows the dynamics for a delay time which is longer than the lifetime of the system. In this case the feedback acts to reintroduce the excitation to the system, in a short pulse, rather than to stabilize or enhance the decay. The phase also becomes much less important and has little effect on the system. This is because the cavity-TLS is essentially in the ground state when the feedback returns; there is no emission for the returning pulse to interfere with.

By replicating these previously studied results -- each addressing its own solution space -- in our QT formalism, we validate the method for the study of feedback effects in cavity-QED systems, at least within the scope of the stated approximations. We have also added extra insight into these regimes by showing example QT graphs and stochastic dynamics.

\subsection{Investigation of the Effect of the  Feedback Loop on Excitation Trapping}
\label{sec:LoopResults}

\begin{figure}[tbp]
    \includegraphics[width=0.499\textwidth]{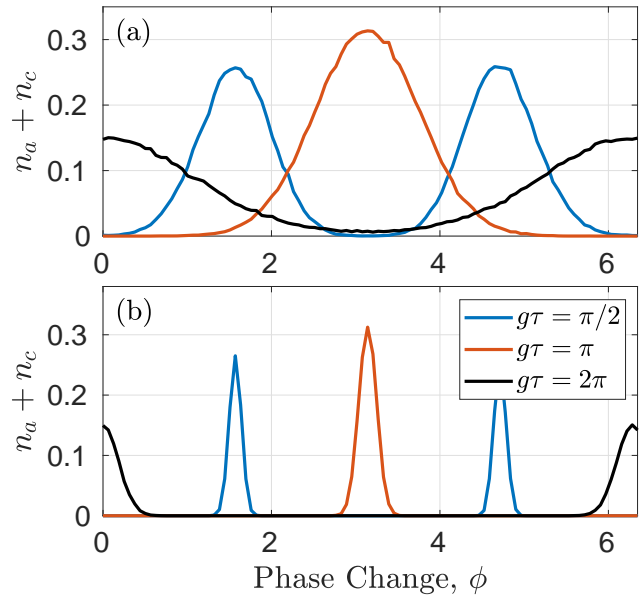}
  \caption{Mean populations of the cavity and TLS at finite $t$, as a function of phase,
  for (a) $t = 30 g^{-1}$ and (b) $t = 1000 g^{-1}$, averaged over 10000 QTs. These QTs were calculated using the optimized technique outlined in Appendix~\ref{sec:App3}. By varying the phase change, $\phi$, of the feedback loop, the optimized phase to improve the lifetime of the system excitation can be found which is different for each loop length. The output to the waveguide for each loop length is $\gamma_{\rm{L}} = 2g$ with everything on resonance, no drive, and no other Lindblad outputs.}
\label{VaryingPhi}
\end{figure}

In the previous section, enhanced spontaneous emission was shown to occur when $\phi = \pi$ and stabilized populations were shown when $\phi = 0$. Which behaviour occurs is not only dependent on the phase, but also the length of the loop, as shown in Fig.~\ref{VaryingPhi}. The peaks in each curve represent the phase at which stabilized populations occur at the chosen loop length. Conversely the troughs of the curve represent regions where enhanced spontaneous emission can be found. Note that the height of the peaks of each curve is dependent on the length of the loop as well. Since the system is an initially excited TLS allowed to decay with no drive, the longer loop lengths essentially `store' population so the stabilized values are lower. Furthermore, the longer feedback time creates more time for the system to decay to the right (and thus out of the system) before the feedback returns.

\begin{figure}[tbp]
    \includegraphics[width=0.499\textwidth]{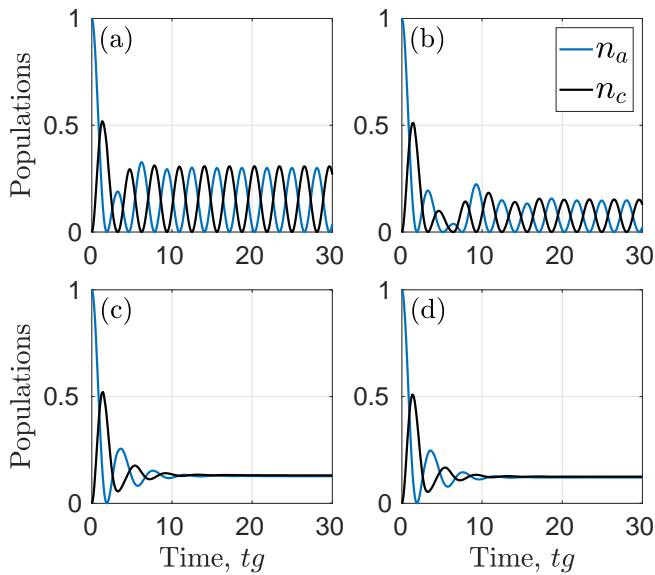}
  \caption{Populations dynamics showing the excitation trapping for different delay times and the required respective phases: (a) $\tau = \pi g^{-1}, \phi = \pi$, (b) $\tau = 2 \pi g^{-1}, \phi = 0$, (c) $\tau = (\pi/2) g^{-1}, \phi = \pi/2$, and (d) $\tau = (\pi/2) g^{-1}, \phi = 3\pi/2$. Each example is an average of 1000 QTs with $\gamma_{\rm{L}} = 2g$, everything is on resonance, and there is no drive or other Lindblad output channels.}
\label{PhiSampleTrajs}
\end{figure}

The location of the peaks in Fig.~\ref{VaryingPhi} is dependent on the phase required to return the reflected field out of phase with the field emitted by the system in order to suppress net emission down the waveguide to the right -- the field ${\cal E}_+(L/2)$ in Fig.~\ref{Model}. When $g\tau = \pi$ or $2 \pi$, this only occurs at one phase, $\phi = \pi$ or $0$ respectively, and leads to stabilized Rabi oscillations as shown in Figs.~\ref{PhiSampleTrajs}(a) and (b). However, when $g\tau = \pi / 2$, this stabilization happens at two different phases, $\phi = \pi / 2$ and $3 \pi /2$. This is because when $\tau = \pi / 2$ there is both a real and imaginary component to $\alpha_1 (t)$ and so the two phases act to match -- and stabilize -- their respective component. However, since only one component can be matched for each $\phi$, the coherence of the Rabi oscillations is lost and a steady state population -- trapped superposition of the TLS and cavity -- is reached as shown in Figs.~\ref{PhiSampleTrajs}(c) and (d). The general condition to achieve excitation trapping, derived in Appendix~\ref{sec:App3}, is given by $\pm g\tau-\phi=2\pi k$ for $k \in \mathbb{Z}$. Note that when there is only one unique solution for $\phi$, as is the case when $g\tau = \pi n$ or $g\tau = 2\pi n$ for $n \in \mathbb{Z}$, then stabilized Rabi oscillations occur. If there is more than one unique solution, then the coherent oscillation is lost and a steady state population is reached. Also, when $\phi$ is moved off of the perfect condition for a trapped excitation, the system excitation will decay away at a rate dependent on how close the parameters are to perfect trapping. In Fig.~\ref{VaryingPhi}(a), the system evolution is truncated at $t = 30 g^{-1}$ while in Fig.~\ref{VaryingPhi}(b) it is truncated after a longer time, at $t = 1000 g^{-1}$, and the width of the peaks decreases while the heights remain the same. If $t \to \infty$, we would be left with a series of delta functions rather than peaks of finite width when we truncate at finite $t$.

\begin{figure}[tbp]
    \includegraphics[width=0.499\textwidth]{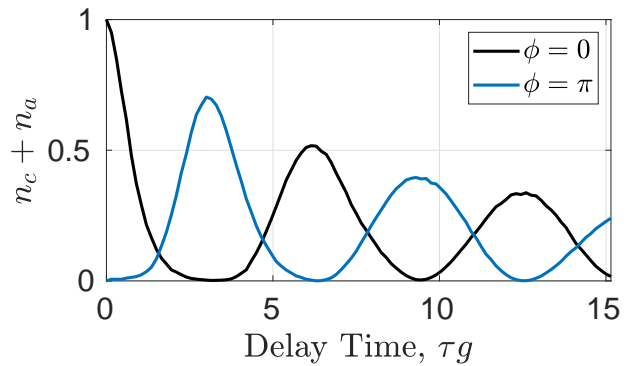}
  \caption{Average populations at time $t = 20 g^{-1}$ for different delay times, $\tau$. As the delay time is longer, less population is held in the cavity-TLS as it spends more time in the loop. The output to the waveguide for each loop length is $\gamma_{\rm{L}} = 0.5g$ with everything on resonance, no drive, and no other Lindblad outputs. Each point is the population in the cavity-TLS (at $t = 20 g^{-1}$) averaged over 10000 QTs generated using the technique outlined in Appendix~\ref{sec:App3}.}
\label{VaryingTau}
\end{figure}

Figure~\ref{VaryingTau} shows this relationship when we fix the phase and allow the delay time to vary. The system oscillates between periods of stabilized Rabi oscillations and enhanced spontaneous emission as the loop length increases. The height of each peak is also decreasing as the delay time grows due to more population being held in the loop rather than in the system and increased time for the system to decay before the feedback returns. As discussed with Fig.~\ref{ShortLong}, when the loop length is increasing the effect of the phase on the system becomes less pronounced. This is seen here through the broadening of each peak as coherence is lost due to the longer round trip time.

\subsection{Nonlinear Cavity-QED Effects}
\label{sec:NonlinearResults}

So far, all of these results have been simulated by setting an initial condition of an excited TLS in vacuum. Since the pump has been turned off, $\Omega = 0$, effectively the system has only had one quanta in it (maximum). Thus the system is essentially linear, and the solution can usually always be solved trivially using frequency-space techniques (e.g., see Ref.~\onlinecite{HughesCoupledQED}) or using an analysis of the delay differential equations (e.g., see Ref.~\onlinecite{Kabuss15}). By turning on the pump beyond a weak excitation, the higher order states of the system can be populated. For example, the $\ket{1} \ket{1} \ket{ \{ 0 \} }$ state (i.e., an excited TLS and one photon in the cavity with no photon in the loop) or the $\ket{1} \ket{0} \ket{ 1,\omega }$ state (i.e., an excited TLS and one photon in the feedback with the cavity in the vacuum state) will be populated. Therefore, by turning on the pump we can begin to look at multi-quanta effects in this system, which cannot be modelled semi-classically, or using the usual weak-excitation approximations. 

\begin{figure}[tbp]
    \includegraphics[width=0.499\textwidth]{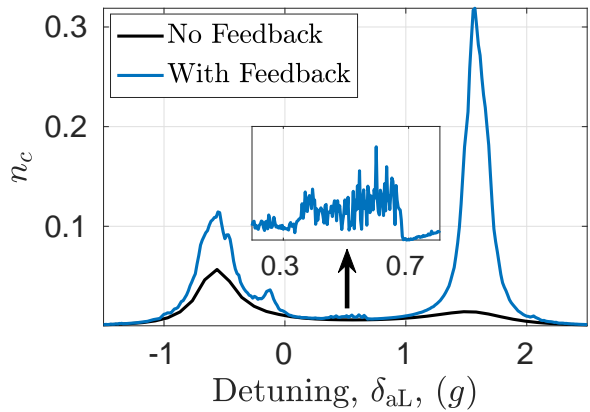}
  \caption{By scanning the detuning from the laser, of strength $\Omega = 0.1g$, the cavity photoluminescence spectrum (proportional to the cavity population) is found by plotting the steady state population of the cavity as a function of detuning. The decay channels have rates $\gamma_{\rm{L}} = 2g, \gamma_{\rm{C}} = 0.05g$, and $\gamma_{\rm{T}} = 0.01g$. The delay time is $\tau = g^{-1}$ and there is no phase introduced by the loop, $\phi = 0$. The detuning between the cavity and TLS is $g$, so that $\delta_{\rm{aL}} = \delta_{\rm{cL}} + g$. Each point is the average of 1500 QTs. The peak introduced by the feedback state is highlighted in the inset at higher detuning resolution.}
\label{Spectra} 
\end{figure}

The system is driven at moderate field strength, $\Omega = 0.1g$, in order to remain within the one photon in the loop approximation while also beginning to see non-linear effects in the system. Figure~\ref{Spectra} shows the cavity photoluminescence spectrum (proportional to the cavity population) of the system with and without feedback when the cavity and TLS have a detuning of $g$. Without feedback, there are only two peaks present, the stronger peak on the left coming from the resonance of the TLS and the weaker peak on the right coming from the cavity resonance. When the feedback is added to the system there is significant enhancement of the two peaks, especially of the cavity resonance due to the feedback returning and stabilizing the cavity population rather than it decaying away.

There are also new peaks that arise from the inclusion of feedback-induced dressed states, which cause additional resonances near $\delta_{\rm{aL}} \sim -0.5g$. These additional peaks are not seen without feedback and are due to the non-linear behaviour introduced and enhanced by the feedback loop. With a stronger pump these non-linear effects will be easier to identify; however, in order to use a stronger pump, higher orders of quanta will need to be allowed in the feedback loop which will be addressed in future work. There are also peaks introduced by the feedback state, $\ket{0}\ket{0}\ket{1,\omega}$ shown in the inset of Fig.~\ref{Spectra}. These peaks occur because of round trip resonances in the feedback loop~\cite{Yao:09}, which appear at $\pm 1/\tau= \pm n g$ (with $n \in \mathbb{Z}$), and we see some signatures of such a retardation peak near $\delta_{\rm{aL}} \sim 0.5g$.

\begin{figure}[tbh]
  \includegraphics[width=0.499\textwidth]{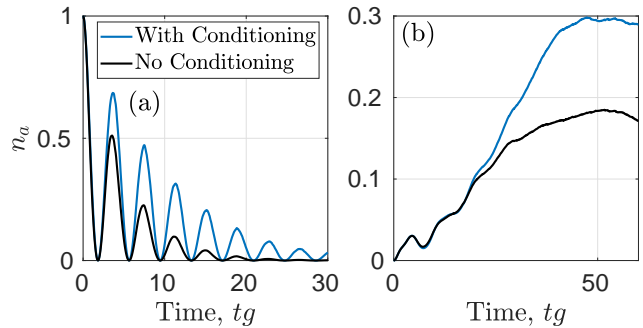}
  \caption{A comparison of the system dynamics with and without conditioning. (a) Decaying Rabi oscillations when $\gamma_{\rm{L}} = 2g$, with no other Lindblad output channels or drive and everything on resonance. The delay time is $\tau = g^{-1}$ without a phase change, $\phi = 0$. (b) The system driven by a weak pump, $\Omega = 0.1g$, with decay rates of $\gamma_{\rm{L}} = 2g$, $\gamma_{\rm{C}} = 0.05g$, $\gamma_{\rm{L}} = 0.01g$, and a detuning of $\delta_{\rm{aL}} = \delta_{\rm{cL}} + g$. The delay time is $\tau = \pi g^{-1}$ and has an overall phase change of $\phi = \pi$. Both figures are an average of 1000 QTs.}
\label{Conditioning}
\end{figure}

It is also important to recognize the importance of correctly conditioning the populations in the presence of the feedback loop. Figure~\ref{Conditioning} compares the technique of conditioning outlined in Sec.~\ref{sec:Conditioning} with the typical renormalization used in Markovian QT theory. In Fig.~\ref{Conditioning}(a) the damping of the Rabi oscillations without conditioning is much faster than when the populations are properly conditioned. Indeed without conditioning it seems as if there is negligible population left in the system at $t = 30 g^{-1}$, but in reality there are still significant Rabi oscillations occurring with the system. When a drive is introduced to the system as in Fig.~\ref{Conditioning}(b), if conditioning is not done properly then not only the incorrect dynamics but also the incorrect steady state population will be found. This is important in order to calculate the correct spectra for the system.

\section{Conclusions}
\label{sec:Conc}

We have presented a QT formalism for simulating the evolution of cavity-QED systems with coherent optical feedback. The equations of motion required to evolve such a system under QT theory are developed and the key quantities required in the numerical simulation are derived. Previous results are recovered using these equations of motion to confirm the accuracy of our derived approach, with QT insights into the stochastic dynamics. Results in the single quanta regime and nonlinear regime are then presented to show the potential of coherent optical feedback to stabilize nonlinear cavity-QED systems and increase their coherent lifetimes. Possible areas of future work for developing this approach with coherent feedback, include extending our model to allow for more than one photon in the loop and computing nonlinear spectra produced from the system.

\acknowledgements
This work was supported by the Natural Sciences and Engineering Research Council of Canada, the Canadian Foundation for Innovation and Queen's University. HJC acknowledges the support of the New Zealand Tertiary Education Committee through the Dodd-Walls Centre for Photonic and Quantum Technologies.

\bibliography{main}

\appendix

\section{Deriving \eqref{FormOfR}}
\label{sec:App1}

We begin deriving \eqref{FormOfR} from the differential equation presented in \eqref{DiffEQ}. It is clear from the structure of $\boldsymbol{A}$ that it is diagonalizable and more importantly that $\boldsymbol{A}$ has four eigenvalues. These eigenvalues are labelled as $\lambda_1, \lambda_2, \lambda_3,\lambda_4$ and their corresponding eigenvectors as $\boldsymbol{\Lambda_1}, \boldsymbol{\Lambda_2}, \boldsymbol{\Lambda_{3}}, \boldsymbol{\Lambda_{4}}$. Then 
\begin{equation}
\boldsymbol{A} \boldsymbol{\Lambda_k} = \lambda_k \boldsymbol{\Lambda_k} = \boldsymbol{\Lambda_k} \lambda_k,  \hspace{0.5cm} k = 1,2,3,4.
\end{equation}

Now define a new matrix $\boldsymbol{E}$ where each column is an eigenvector of $\boldsymbol{A}$,
\begin{equation}
\boldsymbol{E} = [\boldsymbol{\Lambda_1}, \boldsymbol{\Lambda_2}, \boldsymbol{\Lambda_{3}}, \boldsymbol{\Lambda_{4}}],
\end{equation}
so then
\begin{equation}
\boldsymbol{A} \boldsymbol{E} = \boldsymbol{E} \begin{bmatrix}
\lambda_1 & 0 & 0 & 0 \\           
0 & \lambda_2 & 0 & 0 \\
0 & 0 & \lambda_3 & 0 \\
0 & 0 & 0 & \lambda_{4}
\end{bmatrix} = \boldsymbol{E} \boldsymbol{\lambda}.
\label{AERelation}
\end{equation}

The decoupled variables are thus $\boldsymbol{u} = \boldsymbol{E}^{-1} \boldsymbol{R}$ so that $\boldsymbol{R} = \boldsymbol{E} \boldsymbol{u}$. Substituting this into \eqref{DiffEQ} gives
\begin{equation}
\frac{d}{dt} \boldsymbol{E} \boldsymbol{u} = \boldsymbol{A} \boldsymbol{E} \boldsymbol{u} -i G(\omega) \boldsymbol{\alpha\beta},
\end{equation}
and since $\boldsymbol{E}$ is time independent (because all entries of $\boldsymbol{A}$ are time independent) both sides can be multiplied by $\boldsymbol{E}^{-1}$ to get
\begin{equation}
\frac{d}{dt} \boldsymbol{u} = \boldsymbol{E}^{-1} \boldsymbol{A} \boldsymbol{E} \boldsymbol{u} -i G(\omega) \boldsymbol{E}^{-1} \boldsymbol{\alpha\beta}.
\end{equation}
Furthermore, by substituting \eqref{AERelation}, this simplifies to 
\begin{equation}
\frac{d}{dt} \boldsymbol{u} = \boldsymbol{\lambda} \boldsymbol{u} -i G(\omega) \boldsymbol{E}^{-1} \boldsymbol{\alpha\beta}.
\end{equation}
Now since $\boldsymbol{\lambda}$ is a diagonal matrix this can be solved to give an expression for $\boldsymbol{u}$
\begin{equation}
\boldsymbol{u} = -i G(\omega) \int_0^t e^{-\boldsymbol{\lambda}(t' -t)}\boldsymbol{E}^{-1}\boldsymbol{\alpha\beta}dt'.
\end{equation}
Then lastly, we can multiply both sides by $\boldsymbol{E}$ and substitute $\boldsymbol{R} = \boldsymbol{E} \boldsymbol{u}$ to get our final expression for $\boldsymbol{R}$
\begin{equation}
\boldsymbol{R} (\omega,t) = -i G(\omega) \int_0^t \boldsymbol{E}e^{-\boldsymbol{\lambda}(t' -t)}\boldsymbol{E}^{-1}\boldsymbol{\alpha\beta}dt'.
\end{equation}

When the eigenvalues are computed they are all of the form $\lambda_j = -i \omega + c_j$ where $c_j \in \mathbb{C}$ is some constant dependent on the parameters in the Hamiltonian, i.e. $\Omega$, $g$, $\delta_{aL}$, etc. Furthermore, the eigenvectors are frequency independent so they can be written as
\begin{equation}
\boldsymbol{\Lambda_j} = \begin{bmatrix}
           a_{1,j}\\           
           a_{2,j}\\
           a_{3,j} \\
           a_{4,j}
          \end{bmatrix}, \hspace{0.25 cm} a_{i,j} \in \mathbb{C}.
\end{equation}
Then $\boldsymbol{E}$ and $\boldsymbol{E}^{-1}$ are just two matrices of complex numbers 
\begin{equation}
\boldsymbol{E} = [a_{i,j}]_{i,j}, \hspace{0.25cm} \boldsymbol{E}^{-1} = [b_{i,j}]_{i,j}, \hspace{0.25cm} i,j \in \{ 1,2,3,4 \}. 
\end{equation}
Then using this new representation for $\boldsymbol{E}^{-1}$, $\boldsymbol{E}^{-1} \boldsymbol{\alpha\beta} (t')$ is
\begin{equation}
\boldsymbol{E}^{-1} \boldsymbol{\alpha\beta} (t')= \begin{bmatrix}
           b_{1,1}\alpha_1(t') + b_{1,2}\beta_1(t')\\           
           b_{2,1}\alpha_1(t') + b_{2,2}\beta_1(t')\\
           b_{3,1}\alpha_1(t') + b_{3,2}\beta_1(t')\\
           b_{4,1}\alpha_1(t') + b_{4,2}\beta_1(t')
          \end{bmatrix}.
\end{equation}
Also, because $\boldsymbol{\lambda}$ is a diagonal matrix, then $e^{-\boldsymbol{\lambda}(t' -t)}$ is a diagonal matrix as well and using the form of each $\lambda_j$ a factor of $e^{i\omega(t'-t)}$ can be taken out of $e^{-\boldsymbol{\lambda}(t' -t)} \boldsymbol{E}^{-1} \boldsymbol{\alpha\beta} (t')  = e^{i\omega(t'-t)} \boldsymbol{n}(t,t')$ and the vector $\boldsymbol{n}$ is
\begin{equation}
\boldsymbol{n}(t,t') = \begin{bmatrix}
           e^{-c_1(t'-t)}(b_{1,1}\alpha_1(t') + b_{1,2}\beta_1(t'))\\           
           e^{-c_2(t'-t)}(b_{2,1}\alpha_1(t') + b_{2,2}\beta_1(t'))\\
           e^{-c_3(t'-t)}(b_{3,1}\alpha_1(t') + b_{3,2}\beta_1(t'))\\
           e^{-c_4(t'-t)}(b_{4,1}\alpha_1(t') + b_{4,2}\beta_1(t'))
          \end{bmatrix}.
\label{nExpression}
\end{equation}
The important thing to note is that $\boldsymbol{n}(t,t')$ is only a function of $t$ and $t'$ but not $\omega$. Using this expression in the solution for $\boldsymbol{R}(\omega,t)$ gives
\begin{equation}
\boldsymbol{R} (\omega,t) = -i G(\omega) \int_0^t e^{i\omega(t'-t) }\boldsymbol{E} \cdot \boldsymbol{n}(t,t')dt'.
\end{equation}

\section{Deriving \eqref{RIntegral1} and \eqref{RIntegral2}}
\label{sec:App2}

In the equations of motion, the $R_{0,0}(\omega,t)$ and $R_{1,0}(\omega,t)$ terms come in as frequency integrals over all possible frequencies. The form of these two coefficients, shown in \eqref{FormOfR}, already contains a time integral over all past time so we focus on simplifying this double integral for the differential equation of $\alpha_1(t)$, which we call $I$, as both double integrals simplify similarly. Explicitly $I$ has the form 
\begin{equation}
I = - \int_{-\infty}^{\infty} \int_0^t G(\omega')^2 e^{i\omega'(t'-t)} \boldsymbol{E_1}\cdot \boldsymbol{n}(t,t') dt' d\omega'.
\end{equation}
Recalling that $G(\omega) = \sqrt{\gamma_{\rm{L}}/2\pi} \sin[(\omega\tau + \phi)/2]$ and switching the order of integration this becomes
\begin{align}
I ={} & \frac{-\gamma_{\rm{L}}}{2\pi} \int_0^t \boldsymbol{E_1}\cdot \boldsymbol{n}(t,t') \times \\
 & \left[ \int_{-\infty}^{\infty} \sin^2 \left( \frac{\omega\tau + \phi}{2} \right) e^{i\omega'(t'-t)} d\omega'\right] dt'. \nonumber
\end{align}
Substituting
\begin{equation}
\sin[(\omega\tau + \phi)/2] = \left( 1/2i (e^{i(\omega\tau + \phi)/2} - e^{-i(\omega\tau + \phi)/2}) \right)
\end{equation}
into the equation gives the following integral
\begin{align}
I = & \frac{-\gamma_{\rm{L}}}{2\pi} \int_0^t \boldsymbol{E_1}\cdot \boldsymbol{n}(t,t') \left[ \int_{-\infty}^{\infty} \frac{1}{2} e^{-i\omega'(t-t')} \right. \\
 & - \frac{1}{4} \left. \left( e^{-i\omega'(t-t'-\tau)}e^{i\phi} + e^{-i\omega'(t-t'+\tau)} e^{-i\phi} \right) d\omega' \right] dt'. \nonumber
\end{align}
Noting that $\int_{-\infty}^{\infty} e^{-i\omega'X} d\omega' = 2 \pi \delta(X)$ where $\delta(X)$ is the Dirac delta function, the integration of $\omega'$ can be carried out
\begin{align}
I = & \frac{-\gamma_{\rm{L}}}{2\pi} \int_0^t \boldsymbol{E_1}\cdot \boldsymbol{n}(t,t') \left[ \pi \delta(t-t') \right. \\
 & \hspace{1 cm} \left. - \frac{\pi}{2} (\delta(t-t'-\tau)e^{i\phi} + \delta(t-t'+\tau) e^{-i\phi}) \right] dt'. \nonumber
\end{align}
Lastly, noting that $\int_0^t \delta(t''-t') f(t') dt' = f(t'')$ as long as $t'' \in (0,t)$ (or $\int_0^t \delta(t''-t') f(t') dt' = \frac{1}{2}f(t'')$ if $t'' \in \{0,t\}$) the integration of $t'$ can be completed
\begin{align}
-i \int_{-\infty}^{\infty} G(\omega') R_{0,0} (\omega',t) d\omega' & ={} \\
 & \hspace{-3.5 cm} \frac{\gamma_{\rm{L}}}{4} \left( -(\boldsymbol{E_1}\cdot \boldsymbol{n} (t,t)) + e^{i\phi}\theta(t - \tau) (\boldsymbol{E_1}\cdot \boldsymbol{n} (t,t-\tau))\right), \nonumber
\end{align}
where $\theta(t-\tau)$ is the Heaviside step function.

\section{Optimized Technique for Simulating Quantum Trajectories with No Drive}
\label{sec:App3}

The general numerical technique outlined in Sec.~\ref{sec:Numerics} is unnecessarily complex when there is no drive and only one quanta present in the system, which is the case for Secs.~\ref{sec:PrevResults} and \ref{sec:LoopResults}. Since there is only one quanta, it is only possible for one jump to occur during a QT, and after such a jump the system is in the ground state. The only stochastic dynamics that are present in the QTs are when the quantum jumps are chosen to occur, which we can exploit to speed up the computation of our QTs.

By simulating (or solving) the delay differential equation associated with the system, \eqref{CoupledEq2} with \eqref{RIntegral1} and \eqref{RIntegral2} substituted in, and conditioning the result using Sec.~\ref{sec:Conditioning}, the QT without any jumps can be computed. Then each individual QT can be generated by choosing a uniformly distributed random number $\epsilon$, and comparing it to the integrated probability $P(t)$ for a jump to occur. A jump is applied to the system when $\epsilon < P(t)$ and the system collapses to the ground state. Therefore, after the initial QT without jumps is computed, there is no other significant computation to be done for each subsequent trajectory. In the case of Figs.~\ref{VaryingPhi} and \ref{VaryingTau}, each data point is simply the ratio of trajectories where $\epsilon > P(t_{\rm{end}})$ to the total number of trajectories.

Subsequently, by solving the delay differential equation, there are other insights that can be found. In the case of the system parameters for Figs.~\ref{VaryingPhi} and \ref{VaryingTau}, \eqref{CoupledEq2} reduces to
\begin{align}
    \frac{d\beta_0 (t)}{dt} ={} & -ig \alpha_1 (t), \\
    \frac{d\alpha_1 (t)}{dt} ={} & -ig \beta_0 (t) -\frac{\gamma_{\rm{L}}}{4} \alpha_1 (t) + \frac{\gamma_{\rm{L}}}{4} e^{i\phi} \alpha_1(t-\tau). \nonumber
\end{align}
By seeking solutions of the form $\alpha_1(t) = Ae^{i \Lambda t}$ and $\beta_0(t) = Be^{i \Lambda t}$, we arrive at the characteristic equation
\begin{equation}
    -\left\{\Lambda-i\frac{\gamma_L}4[1-e^{i(\phi-\Lambda\tau)}]\right\}\Lambda+g^2=0.
\end{equation}
This equation has the solution $\Lambda = \pm g$ when
\begin{equation}
    \pm g\tau - \phi = 2 \pi k, k \in \mathbb{Z},
\end{equation}
which is precisely the condition we use to find the required phase to trap excitations in Sec.~\ref{sec:LoopResults}.

\end{document}